\documentclass[aps,prx,twocolumn,superscriptaddress,longbibliography]{revtex4-1}

\pdfoutput=1

\usepackage{graphicx}
\usepackage{multirow} 
\usepackage{enumitem} 

\usepackage{hyperref}
\hypersetup{colorlinks=true,urlcolor=blue,citecolor=blue,linkcolor=blue}
\usepackage{cleveref}

\usepackage[caption=false]{subfig} 

\begin{document}

\begin{titlepage}

  \title{Toward more accurate measurement of the impact of online instructional design on students' ability to transfer physics problem-solving skills}
  \author{Kyle M. Whitcomb}
  \affiliation{Department of Physics and Astronomy, University of Pittsburgh, Pittsburgh, PA, 15260}
  \author{Matthew W. Guthrie}
  \affiliation{Department of Physics, University of Connecticut, Storrs, CT, 06269}
  \affiliation{Department of Physics, University of Central Florida, Orlando, FL, 32816}
  \author{Chandralekha Singh}
  \affiliation{Department of Physics and Astronomy, University of Pittsburgh, Pittsburgh, PA, 15260}
  \author{Zhongzhou Chen}
  \affiliation{Department of Physics, University of Central Florida, Orlando, FL, 32816}
  
  \begin{abstract}
    In two earlier studies, we developed a new method to measure students' ability to transfer physics problem solving skills to new contexts using a sequence of online learning modules, and implemented two interventions in the form of additional learning modules designed to improve transfer ability. 
    The current paper introduces a new data analysis scheme that could improve the accuracy of the measurement by accounting for possible differences in students' goal orientation and behavior, as well as revealing the possible mechanism by which one of the two interventions improves transfer ability. 
    Based on a two by two framework of self-regulated learning, students with a performance-avoidance oriented goal are more likely to guess on some of the assessment attempts in order to save time, resulting in an underestimation of the student populations' transfer ability. 
    The current analysis shows that about half of the students had frequent brief initial assessment attempts, and significantly lower correct rates on certain modules, which we think is likely to have originated at least in part from students adopting a performance-avoidance strategy. 
    We then divided the remaining population, for which we can be certain that few students adopted a performance-avoidance strategy, based on whether they interacted with one of the intervention modules designed to develop basic problem solving skills, or passed that module on their first attempt without interacting with the instructional material. 
    By comparing to propensity score matched populations from a previous semester, we found that the improvement in subsequent transfer performance observed in a previous study mainly came from the latter population, suggesting that the intervention served as an effective reminder for students to activate existing skills, but fell short of developing those skills among those who have yet to master it.
  \clearpage
  \end{abstract}

 \maketitle
\end{titlepage}

\section{Introduction}\label{sec:intro}

In addition to learning physics concepts, a key objective of physics instruction is to facilitate students' development of robust problem solving skills and in particular, the ability to transfer the skills that they learned to novel contexts~\cite{bransford1999, broudy1977, detterman1993, Marshman2020}.
How instructional methods can be developed and evaluated to enhance students' transfer ability is a highly valuable research question for STEM education.
However, most existing instruments that assess students' conceptual understanding~\cite{hestenes1992,   thornton1998} or problem solving skills at scale~\cite{pawl2012, marx2010} are not designed to directly measure their ability to transfer, since students were not explicitly provided with the opportunity or the resources to learn and develop new skills during the test. 
Another challenge for accurately assessing students' transfer ability is that the transfer process often involves multiple interleaved stages of learning and problem solving, leading to much richer and more diverse student behavior during the process. 
Yet traditional assessments often lack the ability to provide detailed information on those different student behavior, and how they affect the outcome.
Therefore, it is important to develop new assessment and data analysis methods that can properly capture the complexity of students' behavior during transfer, in order to improve both the accuracy of transfer measurement and our understanding of the mechanism of instructional materials designed to improve transfer.

In an earlier paper~\cite{chen2018perc} we proposed a new method for measuring students' ability to transfer their learning from online problem solving tutorials to new problem contexts by analyzing the log of clickstream data of students interacting with a sequence of online learning modules (OLMs). 
Each module contains both learning materials and assessment problems, as explained in more detail in sections~\ref{subsec:transfer} and~\ref{subsec:olm_sequence}. 
We found that while introductory-level college physics students are highly capable of learning to solve specific problems from online tutorials, they struggled to transfer their learning to a slightly modified problem given immediately afterward on the next module.
In a follow-up study~\cite{chen2019perc}, we tested two different methods to enhance students' ability to transfer in an OLM sequence and found evidence suggesting that the addition of an ``on-ramp'' module (a scaffolding module designed to solidify essential basic skills and concepts~\cite{mikula2017, young2018}) prior to the tutorial resulted in significant improvement in students' ability to transfer their knowledge in the rotational kinematics sequence, while the second intervention did not result in significant differences in the outcome.

The design of the OLM modules enabled multiple levels of transfer to take place by integrating the instruction and assessment, but our initial analysis did not examine whether students interacted with those modules as we had intended, nor did our previous analysis verify the mechanism by which the ``on-ramp'' module improved transfer. 
Therefore, the current study will improve the quality of analysis by answering the following research questions. 
First, since the OLMs are assigned for students to complete on their own, what fraction of students interacted with the modules as we had intended? For those who did not, to what extent did their alternative strategy, as described in section \ref{subsec:learning_strategies}, affect the validity of our measurement of students' transfer ability, and how can we mitigate those impacts for a more accurate measurement?
Second, as earlier analyses suggested that the ``on-ramp'' modules may be effective, what is the mechanism by which those modules enhance students' transfer performance in subsequent modules?
Are the benefits of those modules exclusive only to students who interacted with them in a certain way, as explained in section \ref{subsec:onramp_mechanisms}?

\subsection{Measuring transfer in an OLM sequence}\label{subsec:transfer}
As will be explained in more detail in section \ref{sec:methods}, each OLM consists of an instructional component (IC) and an assessment component (AC) which contains one or two problems, as demonstrated in Fig.~\ref{fig:design} adapted from Ref.~\cite{chen2018perc}.
Students are required to complete at least one attempt on the AC before being allowed to study the IC, a design that was inspired by the frameworks of preparation for future learning~\cite{bransford1999} and productive failure~\cite{kapur2009}.
Students who failed their first attempt can learn to solve the specific type of problem from the IC. 
When students complete a sequence of two or more OLMs in sequence on the same topic involving similar assessment problems, their required first attempt on the subsequent module serves as an assessment of their ability to transfer their learning from the IC of the previous module.
When more than two modules are involved, students' performance on later modules could be attributed to indirect transfer due to a preparation for future learning effect; that is, completing the first module better prepares students to learn from the second module, which in turn increases performance on the third and subsequent modules.

\begin{figure}
  \includegraphics[width=\linewidth]{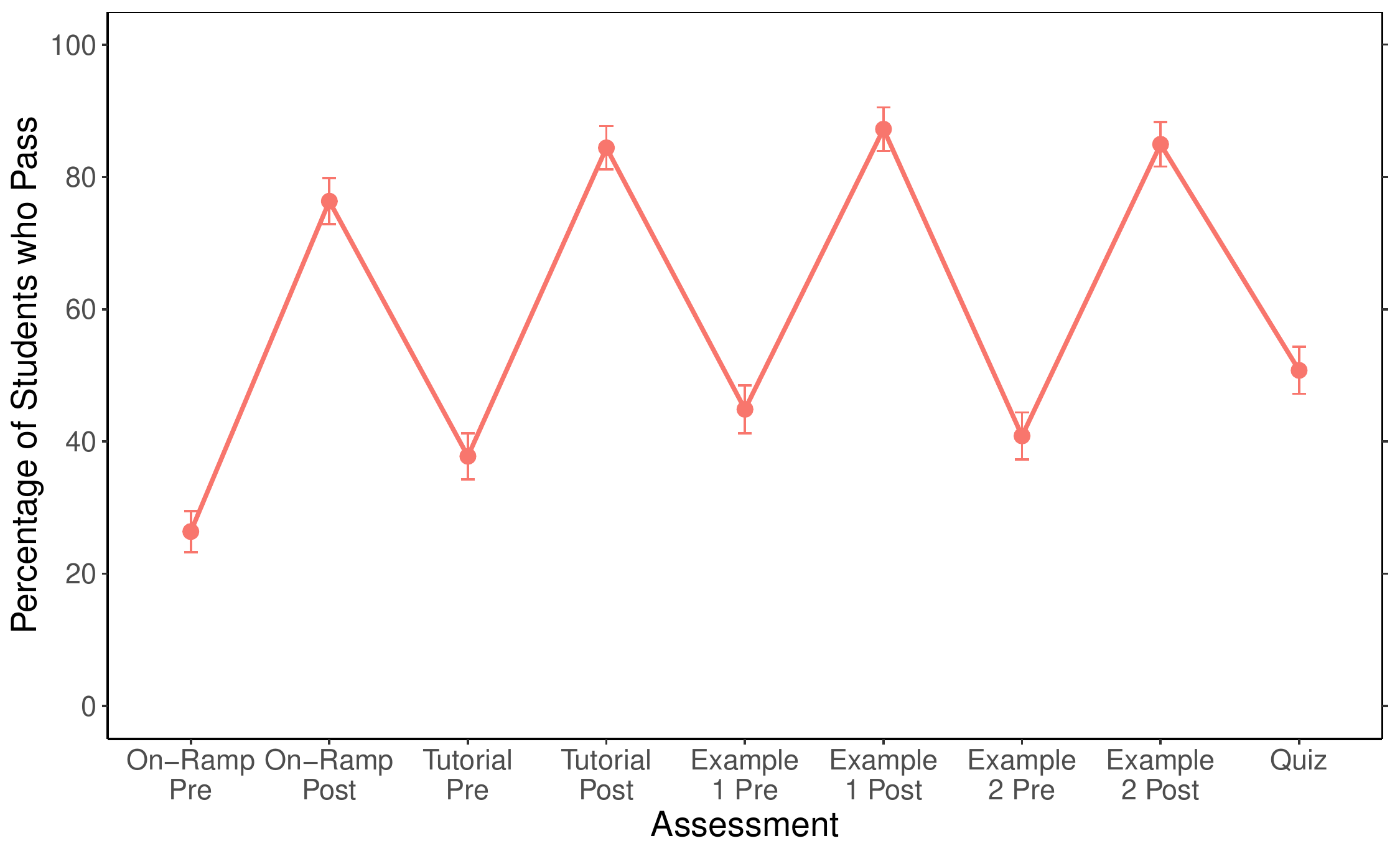} 
  \caption{
  An example of a zigzag plot, adapted from Ref.~\cite{chen2019perc}.
  Each point represents the passing rate of students either before (``Pre'') or after (``Post'') being given access to the instructional material in each module.
  Passing rates in the Post-stage of a module are cumulative with Pre-stage attempts.
  See section \ref{subsec:analysis} for more details.
  }
  
  \label{fig:zigzag_example}
\end{figure}

Data from OLMs can be visualized in a ``zigzag'' plot (Fig.~\ref{fig:zigzag_example}, adapted from Ref.~\cite{chen2019perc}), developed in earlier studies and explained in detail in section \ref{subsec:analysis}.
Every two points represent the total assessment passing percentage of the student population on attempts before and after learning from the IC of each module.
Students' ability to learn to solve a specific problem is reflected by an increase in passing percentage from Pre to Post on the same module.
The odd-numbered points in Fig.~\ref{fig:zigzag_example} (i.e, those labeled ``Pre'' as well as ``Quiz'') show passing rates on initial attempts prior to learning from the IC of each module, and an increase from one point to the next reflects students' ability to transfer their learning from the previous module(s).

\subsection{Students' different learning strategies and possible impact on assessment}\label{subsec:learning_strategies}

Measuring students' transfer ability from their performance on OLM assessment attempts requires that the majority of students either seriously took the required first attempt of each module, or made a quick guess only when they feel that they cannot solve the problem. 
However, research on students' self-regulated learning (SRL) processes suggests that learners may choose to guess regardless of their ability or confidence to solve an assessment problem according to their motivational goal orientation. 
Using a $2 \times 2$ achievement goal framework~\cite{Elliot2008, Elliot2001}, learners' goals can be classified along both the definition dimension and the valence dimension. 
On the \textit{definition} dimension, the learner can be either mastery-oriented or performance-oriented. 
Simply put, mastery-oriented learners focus more on and are mostly motivated by the intrinsic value of mastering the subject, while performance-oriented learners are motivated by extrinsic values (see also the summary of Pintrich's model~\cite{pintrich2000,pintrich2004} by Winne~\cite{winne2015}), such as obtaining the homework credit for each module. 
On the \textit{valence} dimension, learners either focus on a ``positive possibility to approach (i.e., success)'' or on a ``negative possibility to avoid (i.e., failure).'' 

It is easy to imagine that if a learner has a performance-avoidance type achievement goal, then they are likely to adopt a strategy akin to a ``coping mode,'' described by Boekaerts~\cite{boekaerts2000} as primarily focusing on ``preserving [study] resources and avoiding damage.'' 
In the context of interacting with OLM modules, a student with a \textit{performance-avoidance goal} is likely to randomly submit an answer on their required first attempt to avoid ``unnecessary failure'' and save time, and then study the IC to ensure success on their next attempt.
For those students, their initial attempts reflect their learning strategy, rather than their level of content mastery, transfer ability, or even their confidence.

If some students in our sample did adopt such a strategy, then the log data of their interactions with the modules will have two characteristic features:
1) their initial attempts will frequently be significantly shorter in time and have much lower passing rates when compared to other students, at least on some of the easier modules; 2) their passing rate on attempts after study will be similar to everyone else.

If a non-negligible fraction of students indeed adopted the performance-avoidance strategy, their data could significantly distort the estimation of transfer ability for the entire student population.
Properly identifying and removing those students from the sample will improve the accuracy of the measurement using data from OLMs.

\subsection{Distinguishing between two different mechanisms of the on-ramp module}\label{subsec:onramp_mechanisms}
In our earlier study~\cite{chen2019perc}, we found that the addition of an ``on-ramp'' module at the beginning of the OLM sequence resulted in better performance on the required first attempts for subsequent modules compared to students from the previous semester. 
The ``on-ramp'' modules contain practice problems designed to develop and enhance students' proficiency of essential skills necessary for problem solving.
However, students who passed the AC of the on-ramp module on their required first attempt (or on attempts before accessing the IC) can choose to directly move on to the next module without interacting with the IC of the on-ramp module.
Therefore, if the on-ramp module enhances students' transfer ability by improving their proficiency on essential skills, then the improvement will not be statistically significant among those who passed on the first attempt, and statistically significant among those who failed their initial attempt and accessed the IC.
Alternatively, if the on-ramp module mainly serves as a ``reminder'' for students to activate existing knowledge of essential skills, then the benefit should be more significant among those who passed on the first attempt, and much smaller for those who studied the IC.
Distinguishing between those two mechanisms can better guide the future development of instructional materials to enhance students' ability to transfer.

\subsection{Research questions}\label{subsec:rq}
To summarize, in this study we will answer the following three research questions:

\begin{enumerate}[label={\bfseries RQ\arabic*}, topsep=5pt, itemsep=0pt]
  \item \label{rq:behavior} What fraction of students display the characteristic features in the log data that is indicative of adopting a performance-avoidance strategy when interacting with OLM sequences?
  \item \label{rq:engagement}
  If we assume that a significant portion of students who display the characteristic features of a performance-avoidance strategy did adopt that strategy, how would the results of previous studies change if we restrict the study to those students who did not display those features?
  \item \label{rq:onramp} Did the on-ramp module enhance students' ability to transfer by improving students' proficiency in essential skills or by serving as a ``reminder'' for those who are already proficient?
\end{enumerate}

The first two research questions are important for the accuracy of the measurements, and lay the groundwork for answering \ref{rq:onramp}. 
In sections \ref{subsec:olm_sequence} to \ref{subsec:data}, we will explain in detail the structure and implementation of OLM sequence, as well as the data collection process. 
In section \ref{subsec:analysis}, we present our operational definition of key concepts such as assessment passing percentage and performance-avoidance strategy in the context of OLMs and outline our analysis procedure for measuring transfer and answering the research questions. 
In section \ref{sec:results}, we present the results of our analysis, which are interpreted in section \ref{subsec:interpretation-of-results}, and their implications are discussed in the rest of \ref{sec:discussion}.

\section{Methods}\label{sec:methods}

\subsection{OLM Sequence Structure}\label{subsec:olm_sequence}

\begin{figure}
  \centering
  \includegraphics{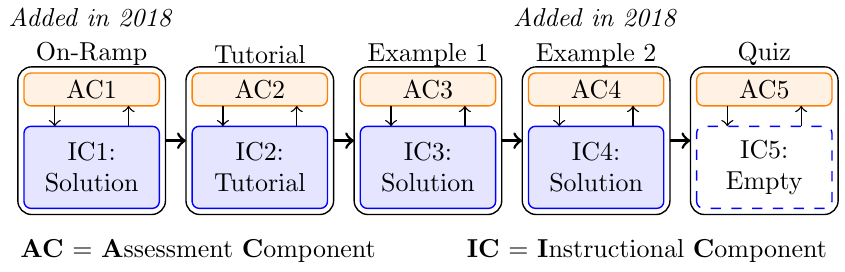}
  
  \caption{
  The sequence of Online Learning Modules (OLMs) designed for this experiment.
  Each OLM contains an assessment component (AC) and an instructional component (IC).
  Students are required to make at least one attempt on the AC first, then are allowed to view the IC, and go on to make subsequent attempts on the AC.
  OLMs 1 and 4 were added for the 2018 implementation.
  }
  \label{fig:design}
\end{figure}

The study was conducted using online learning modules (OLMs)~\cite{chen2017perc, chen2018re, chen2018perc, chen2019perc} implemented on the open source Obojobo platform~\cite{obojobo} developed by the Center for Distributed Learning at the University of Central Florida (UCF).
Each OLM contains an assessment component (AC) and an instructional component (IC) (see Fig.~\ref{fig:design}.
Students have 5 attempts on the AC, which contains 1-2 multiple-choice problems, and must make at least one attempt before being allowed to access the IC.
The IC contains instructional text, figures, and/or practice questions in general.
Specific contents of the IC used in each of the modules in the current study will be detailed in the next section.
In an OLM sequence, a student must either pass or use up all five attempts on the AC before being allowed to access the next module.
Students' interaction with each OLM can be divided into three stages: The pre-study (Pre) stage in which a student makes one or more attempts on the AC, the study stage in which those who failed in the Pre stage study the IC, and the post-study (Post) stage in which students make additional attempts on the AC.
A small fraction (approximately 10\%) of students have also been observed to choose to skip the study stage after more than 3  failed attempts in the Pre stage.
A student is counted as passing an AC if the student correctly answers all problems in the AC within their first 3 attempts, including both Pre and Post stage attempts.
In other words, students who either failed on all 5 attempts or passed on their 4th or 5th attempts are considered as failing the module in the current study.
Because students who skipped the study stage after 3 or more failed attempts will always be categorized as ``Fail,'' the fact that they never accessed the instructional material will not impact any of the analysis in the current study. 

\subsection{Study Setup}\label{subsec:study_setup}

In Fall 2017, two sequences each containing 3 OLMs (specifically, OLMs 2, 3, and 5 in Fig.~\ref{fig:design}) were assigned as homework to 235 students enrolled in a calculus-based introductory physics class at UCF~\cite{chen2018perc}.
The 6 modules were worth 3\% of the total course credit.
The first OLM sequence teaches students to solve Atwood machine type problems with blocks hanging from massive pulleys using knowledge of rotational kinematics (RK).
The second sequence teaches students to solve angular collision problems such as a girl jumping onto a merry-go-round using knowledge of conservation of angular momentum (AM).
Both sequences are designed to develop and measure students' ability to transfer problem solving skills to slightly different contexts. 
The modules used in this study are free and available to the public at Ref.~\cite{obojobo-sample}.

The AC of each OLM contains one problem that can be solved using the same physics principles as other ACs in the OLM sequence.
The IC of OLM 2 (Fig.~\ref{fig:design}) contains an online tutorial developed by DeVore and Singh~\cite{devore2017, singh2010}, in the form of a sequence of practice questions.
The IC of OLM 3 contains a worked solution to the AC problem, and the IC of OLM 5 is empty since it is intended to serve the role of a quiz.

In Fall 2018, the two OLM sequences were each modified by adding two additional OLMs (shown in Fig.~\ref{fig:design}) and implemented again in the same course taught by the same instructor as homework to 241 students.
Both sequences were assigned as homework that was worth 3\% of the total course credit.
The first new module in each sequence is the ``on-ramp'' module (OLM 1 in Fig.~\ref{fig:design}), which contains an AC focusing on one or more basic procedural skills necessary for solving the subsequent ACs in the OLM sequence.
For the RK sequence, the on-ramp module presents students with two Atwood machine problems of the simplest form, involving one or two blocks hanging at the same radius from a single massive pulley.
For the AM sequence, the on-ramp module addressed the common student difficulty of calculating both the magnitude and sign of the angular momentum of an object traveling in a straight line about a fixed point in space.
The second new module in each sequence is the ``Example 2'' module (OLM 4 in Fig.~\ref{fig:design}), which contains in its AC a new problem that shares the same deep structure as the one in the previous module, but differs in surface features.
The IC of the module was designed in two formats: a compare-contrast format in which students were given questions that prompted them to compare the similarity and difficulty of the solutions to the problems in AC3 and AC4, and a guided tutorial format consisting of a series of tutorial-style scaffolding questions guiding them through the solution of the problem in AC4.
Each form was provided to half of the student population at random.
We found no difference between the two cohorts in terms of students' behavior and performance on the subsequent module 5~\cite{chen2019perc}.

\subsection{Data Collection and Selection}\label{subsec:data}

Anonymized clickstream data were collected from the Obojobo platform for all students who interacted with the OLM sequences.
The following types of information were extracted from the log data following the same procedure explained in detail in Ref.~\cite{chen2020}: the number of attempts on the AC of each module, the outcome of each attempt (pass/fail), the start time and duration of each attempt, and the start times of interaction with the IC.
The duration of interaction with the IC was also extracted but was not used in the current analysis.

In addition, students' exam scores and overall course grades, each on a 0-100 scale, were also collected, anonymized, and linked to each students' log data.
The exam scores consist of two midterm exams, each counting for 12\% of the final course grade, and a final exam counting for 16\% of the final course grade.
The final course grade also contains scores from homework, lab, and classroom participation.

In order to maintain a consistent sample across our analyses, only data from students who attempted every module in a sequence at least once are included.
Data from seven students for the 2017 RK sequence were removed because of this reason, and two or fewer students were removed for all other OLM sequences.
Data from 202 students were retained for the RK sequence in 2017, 198 students in the RK sequence for 2018, 198 students for the AM sequence in 2017, and 189 students for the AM sequence in 2018.

In the Fall 2017 implementation, half of the students were given the option to skip the initial AC attempt of OLM 2 (the first OLM in that implementation) and proceed directly to the tutorial in the IC.
However, we found in an earlier study~\cite{chen2018perc} that very few students chose to exercise this option and among those who did there was no detectable impact on subsequent problem solving behavior and outcome.
Therefore, in the current analysis, we combined those two groups into one.
Similarly, for the Fall 2018 semester, we combined data from students encountering the two different versions of IC in module 4, since no difference in their behavior and outcome on module 5 could be detected~\cite{chen2019perc}.

\subsection{Data Analysis}\label{subsec:analysis}

To estimate the fraction of students adopting a performance-avoidance strategy (\ref{rq:behavior}), we will analyze the frequency of students making a very brief first attempt on each module.
As explained in section~\ref{subsec:learning_strategies}, students who adopt such a strategy are more likely to consistently guess on their first attempts and gain access to the instructional material.

In the current analysis, we categorize each student's first attempt as a ``Brief Attempt'' (BA) if the duration of the attempt is less than 35 seconds.
This cutoff time is inherited from a careful analysis of similar OLMs in an earlier study~\cite{chen2020}, and chosen as a conservative estimate for the minimum amount of time needed to read and submit an answer to a given question.
Students are categorized into three ``BA groups'' based on the number of BAs on the first four modules: 0-1 BAs, 2-3 BAs, and 4 BAs.
Table~\ref{tab:2018_BA_groups} shows the number of students in each BA group for each OLM sequence.
BAs on the quiz module were not considered since there was no IC for the students to access.
The 0-1 BA group is the one with the fewest performance-avoidance focused students, and are most likely to make valid first attempts on the AC, whereas students in the 4 BA group are most likely to adopt such a strategy. 

\begin{table}[t]
  \caption{
  The number of students in each OLM sequence by their number of Brief Attempts.
  The Brief Attempt groups consist of those who had 0-1, 2-3, or 4 Brief Attempts throughout the first four modules.}
  \label{tab:2018_BA_groups}
  \centering
  \begin{ruledtabular}
  \begin{tabular}{r | c c c}
      OLM     & \multicolumn{3}{c}{\# of Brief Attempts} \\
      Sequence  & 0-1  & 2-3  & 4 \\
      \hline
      RK     & 100  & 82  & 16 \\
      AM     & 91  & 71  & 27
  \end{tabular}
  \end{ruledtabular}
\end{table}

\begin{figure*}
  \subfloat[Rotational Kinematics\label{fig:zigzag_BA:RK}]{\includegraphics[width=0.49\linewidth]{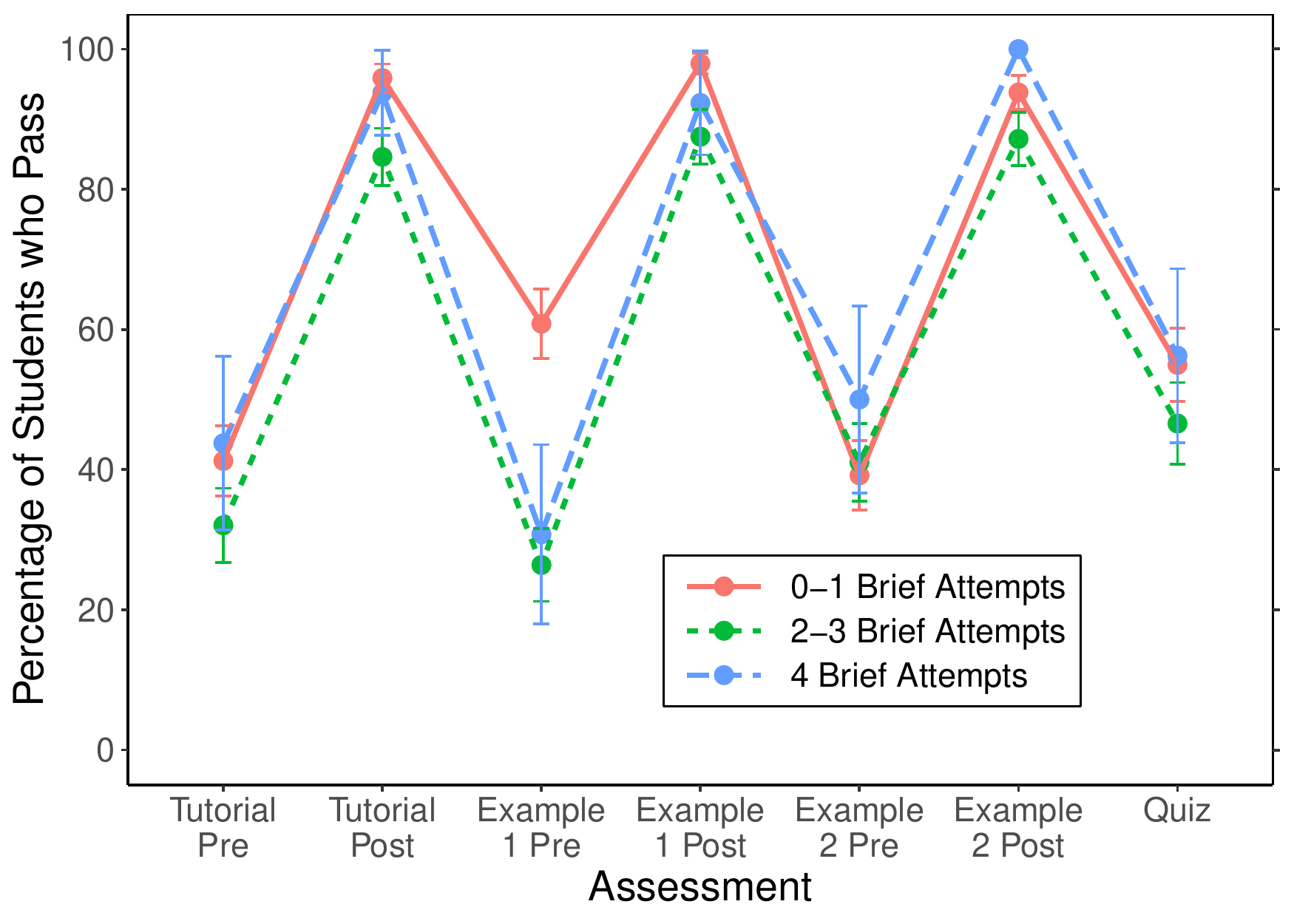}}%
  \hfill
  \subfloat[Angular Momentum\label{fig:zigzag_BA:AM}]{\includegraphics[width=0.49\linewidth]{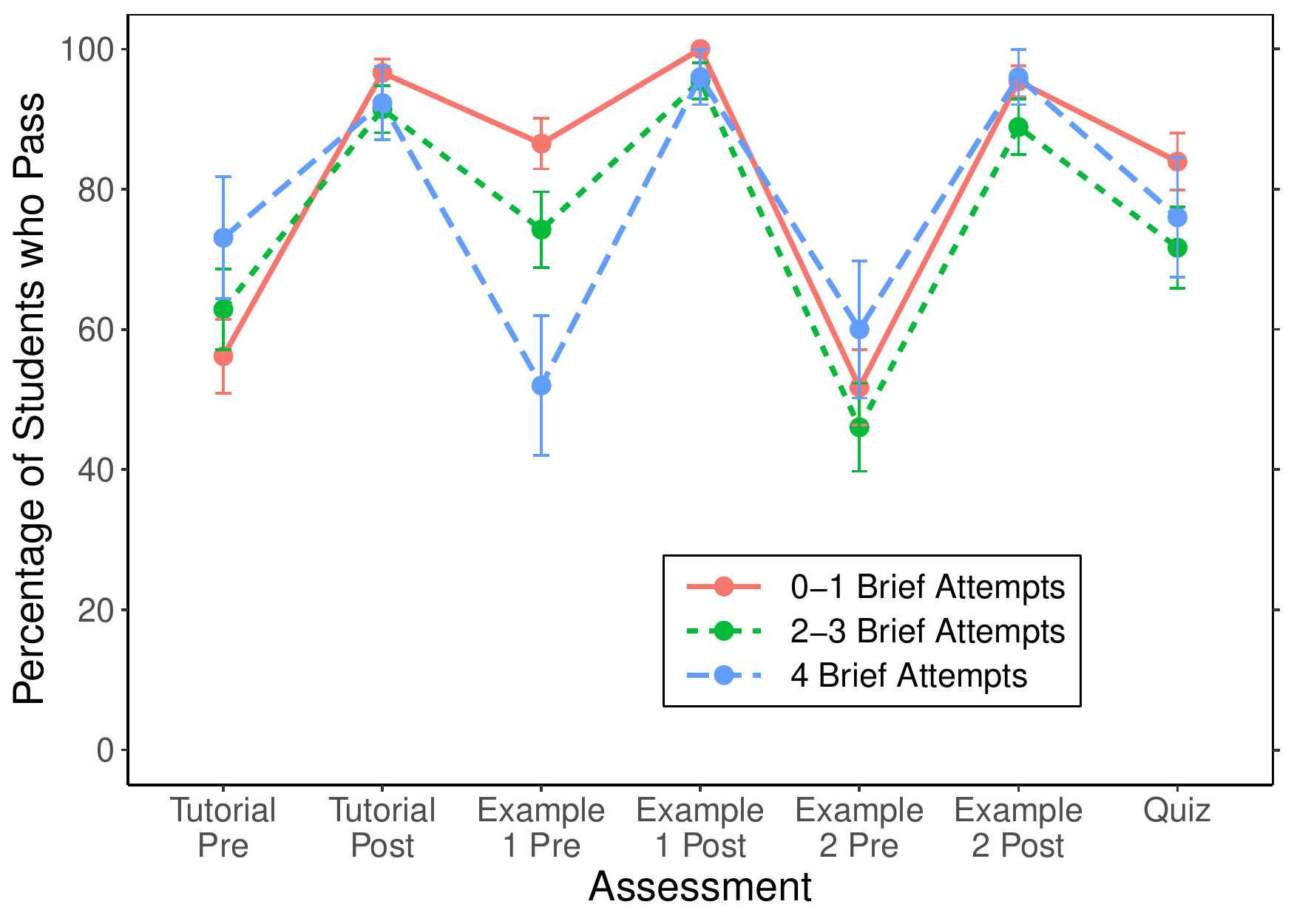}}
  \caption{
  Comparison of performance on OLMs between students with different numbers of Brief Attempts (a) Rotational Kinematics and (b) Angular Momentum.
  The error bars represent standard error. Passing rates on the on-ramp module is not shown since it is irrelevant to the discussion of transfer.
  }
  
  \label{fig:zigzag_BA}
\end{figure*}

To examine the extent to which the behavior of performance-avoidance focused students affect the measurement of transfer (\ref{rq:engagement}), we will compare the Pre and Post stage passing rates of the three BA groups on all modules in the two sequences, and plot the outcomes in Fig.~\ref{fig:zigzag_BA}.
Following the convention established in two previous studies~\cite{chen2018perc, chen2019perc}, the pass rates are defined as follows.
On each OLM module except for module 5, the pass rates ($P$) of students was calculated for both the Pre-study ($P_{\textrm{pre}}$) and Post-study attempts ($P_{\textrm{post}}$).
The Pre-study pass rate on each module is calculated as
\begin{equation}
  P_\textrm{pre} = \frac{N_\textrm{pre}}{N_\textrm{total}},
\end{equation}
with $N_\textrm{pre}$ being the number of students who passed Pre-study and $N_\textrm{total}$ being the total number of students who attempted the module.
Similarly, the Post-study pass rate on each module is calculated as
\begin{equation}
  P_\textrm{post} = \frac{N_\textrm{pre} + N_\textrm{post}}{N_\textrm{total}},
\end{equation}
with $N_\textrm{post}$ being the number of students who passed Post-study.
By including both $N_\textrm{pre}$ and $N_\textrm{post}$, the Post passing rate reflects the total number of students able to pass the assessment after being given the access to the IC, assuming that students who passed in the Pre stage can also pass in the Post stage if re-tested. This definition is similar to the Post test score in a Pre-test/Post-test setting.
For module 5, the passing rate does not distinguish between Pre and Post stage because the IC of the module contains no instructional resources. The $P_{\textrm{pre}}$ on modules 2-4 and $P$ on module 5 measures students' ability to transfer their learning from modules 1-4.
We hypothesized that the 0-1 BA group would have significantly better performance than the other two BA groups on their Pre stage attempts on modules 2, 3, and 4 because the other two BA groups are more likely to forfeit the first attempt opportunity regardless of their ability to solve the problem.
We further hypothesized that the Post-study pass rates for each BA group will be very similar, because $P_\mathrm{post}$ reflects students ability to learn from the modules and solve the specific problem (if they are not already proficient), and the dominant factor separating the three groups is students' engagement strategy, not their ability to learn from the modules.

Finally, to examine the mechanism by which the on-ramp module improves transfer of knowledge (\ref{rq:onramp}), we first separate the student sample from Fall 2018 into three ``on-ramp cohorts'':
\begin{itemize}
  \item \textbf{Pass On-Ramp Pre}: students who passed the on-ramp AC before accessing the IC,
  \item \textbf{Pass On-Ramp Post}: students who passed the on-ramp AC only after accessing the IC, and
  \item \textbf{Fail}: students who did not pass the on-ramp AC within 3 attempts.
\end{itemize}
For this analysis, only data from the 0-1 BA group will be retained. As will be discussed in more detail in sections \ref{sec:results} and \ref{sec:discussion}, the analysis of \ref{rq:behavior} and \ref{rq:engagement} suggests that students in the other two BA groups indeed displayed the characteristic features of a performance-avoidance strategy and thus are much more likely to have adopted such a strategy.
Therefore, it is possible that including those students will which could result in an underestimation of students' ability to transfer.
To that end, the following analysis method will produce accurate results for students in the 0-1 BA group only.

\begin{figure*}
  \subfloat[Rotational Kinematics\label{fig:zigzag_matched_all:RK}]{\includegraphics[width=0.49\linewidth]{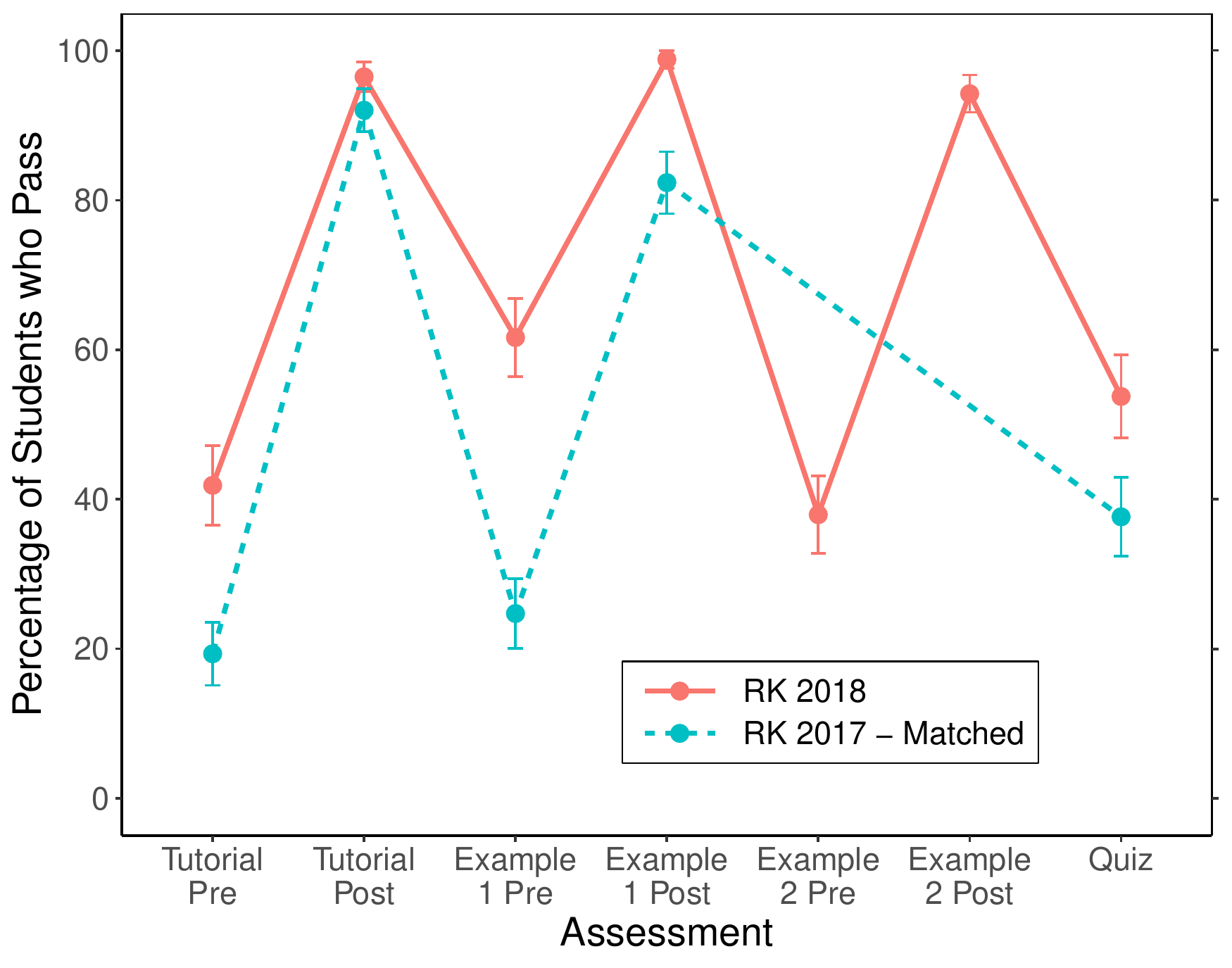}}%
  \hfill
  \subfloat[Angular Momentum\label{fig:zigzag_matched_all:AM}]{\includegraphics[width=0.49\linewidth]{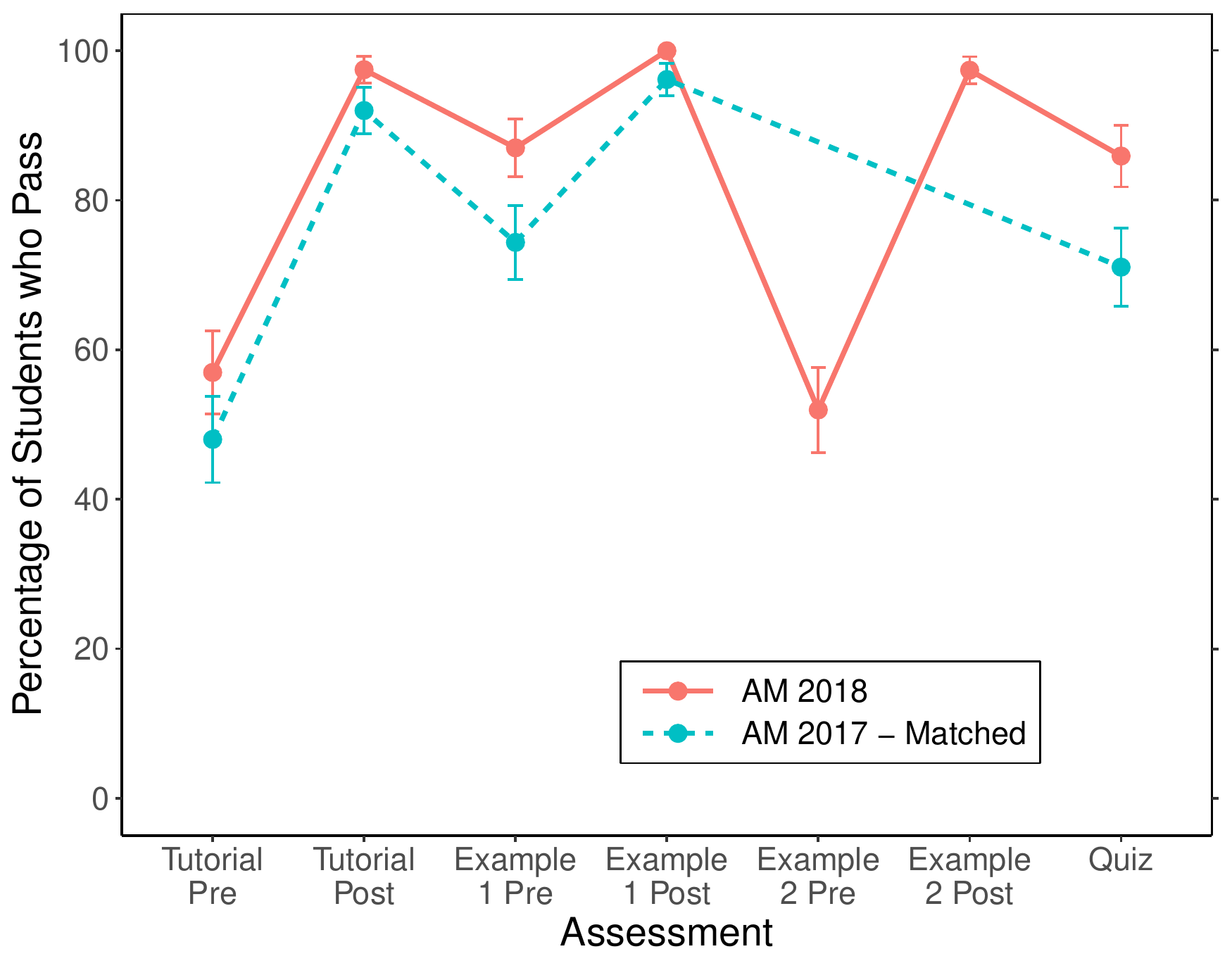}}
  \caption{
  Comparison of the performance on the pre and post attempts of module for students with 0-1 Brief-attempts. 
  A subset of 2017 students were selected to match the background knowledge of 2018 students using a propensity score derived from exam scores. Passing rates on the on-ramp module is not shown. Data on Example 2 from 2017 is absent because the module was added in 2018.
  }
  \label{fig:zigzag_matched_all}
\end{figure*}

Next, we identified three comparable cohorts of students from the 2017 sample.
We first retained students who only made 0-1 BA on modules in the 2017 sequence, then identified comparable cohorts using propensity score matching, since the general ability of the 0-1 BA group could be different from the rest of the student population.
Propensity scores were constructed using a combination of standardized scores from two mid-term exams and one final exam in both semesters.
Each exam is largely identical across the two semesters, with one or two questions being replaced or modified.

Pass rates on modules 2-5 in both sequences are compared between the three 2018 cohorts and the three propensity score matched 2017 cohorts in order to distinguish between the two possible mechanisms for of the on-ramp module.
If the ``improve proficiency'' effect was dominant, then the performance differences should be observed mostly among the Pass On-Ramp Post cohort and its matched cohort in 2017. If the ``reminder'' effect was dominant, then the differences will be observed for the Pass On-Ramp Pre cohort and its counterparts.

Propensity score matching was performed using R~\cite{rcran} and the \texttt{MatchIt} package~\cite{matchit_package}.
The MatchIt algorithm retains all treated data and attempts to find either an exact one-to-one match or balance the overall covariant distribution for the control data.
As shown in Table \ref{tab:psm}, the matching program reduced the difference in the mean of the normalized propensity score in every case. 

\begin{table}[b]
  \caption{The mean difference in propensity scores between the listed 2017 and 2018 on-ramp cohorts both before and after propensity score matching was carried out.
  All students in these samples are in the 0-1 BA group.}
  \label{tab:psm}
  \centering
  \begin{ruledtabular}
  \begin{tabular}{l l l l}
      OLM       &                   & Mean Difference   & Mean Difference \\
      Sequence  & On-Ramp Cohort    & Before Matching   & After Matching \\
      \hline
      RK        & All               & 0.0272      & 0.0083 \\
      RK        & Pass On-Ramp Pre  & 0.0061      & 0.0003 \\
      RK        & Pass On-Ramp Post & 0.0410      & 0.0044 \\
      AM        & All               & 0.0388      & 0.0105 \\
      AM        & Pass On-Ramp Pre  & 0.0599      & 0.0126 \\
      AM        & Pass On-Ramp Post & 0.0286      & 0.0001
  \end{tabular}
  \end{ruledtabular}
\end{table}

Data analysis, statistical testing, and visual analysis were conducted using R~\cite{rcran} and the \texttt{tidyverse} package~\cite{tidyverse}.

\section{Results}\label{sec:results}


First, we measure the fraction of students that displayed characteristic features in their activity log indicative of a performance-avoidance strategy (\ref{rq:behavior}).
We start by listing the number of students with 0-1, 2-3, or 4 BAs on the first four modules of each sequence in Table~\ref{tab:2018_BA_groups}.
The result shows that, even with relatively conservative criteria for classifying brief attempts, we still identified 10-15\% of the students who made four brief attempts the four modules (4 BA group).
On the other hand, around 50\% of the students belong to the 0-1 BA group.

\begin{table}[b]
  \caption{The number of students in each OLM sequence that fall into each on-ramp cohort among those with 0-1 Brief Attempts.
  The cohorts consist of those who passed during on-ramp Pre-study attempts (``Pass On-Ramp Pre''), those who passed during on-ramp Post-study attempts (``Pass On-Ramp Post''), and those that failed the on-ramp assessment (``Fail'').
  Since the on-ramp module was only included in Fall 2018, only students from 2018 are included here.}
  \label{tab:2018_onramp_groups}
  \centering
  \begin{ruledtabular}
  \begin{tabular}{r | c c c}
      OLM     & Pass On-Ramp & Pass On-Ramp  & \multirow{2}{*}{Fail} \\
      Sequence  & Pre   & Post  & \\
      \hline
      RK     & 32      & 57      & 11 \\
      AM     & 32      & 47      & 12
  \end{tabular}
  \end{ruledtabular}
\end{table}

Figure~\ref{fig:zigzag_BA} shows the Pre and Post stage pass rates of students on modules 2-5, separated by BA groups.
Pass rates from the two sequences are plotted separately: the RK sequence in Fig.~\ref{fig:zigzag_BA:RK} and the AM sequence in Fig.~\ref{fig:zigzag_BA:AM}.
In both Fig.~\ref{fig:zigzag_BA:RK} and Fig.~\ref{fig:zigzag_BA:AM}, the most prominent difference between the three BA groups is that students in the 0-1 BA group significantly outperformed the other two groups in Pre stage attempts for the Example 1 module (OLM 2, Fig~\ref{fig:design}) (Fisher's exact test on $2 \times 3$ contingency tables, $p<0.001$ for the RK sequence and $p=0.001$ for the AM sequence).
Students in the 0-1 BA group also outperformed the 2-3 BA group on RK Tutorial Post Stage attempts ($p=0.028$) and RK Example 1 Post stage attempts ($p=0.018$), but the those two groups didn't show a statistically significant and consistent difference with the 4 BA group. None of the other data points showed significant differences between the three groups.


The observations that the 0-1 BA group significantly outperforms the 2-3 BA and 4 BA groups on the Pre stage attempts on the Example 1 module and that the performance differences are much smaller on most of the post-study attempts fits the description of students adopting a performance-avoidance strategy, described in section~\ref{subsec:learning_strategies} and discussed further in section~\ref{subsec:interpretation-of-results}.
Therefore, it is reasonable to assume that at least some of the students in those two groups had adopted a performance-avoidance strategy to some extent.
If we accept this assumption, then the statistically significant performance differences on Example 1 (Fig.~\ref{fig:zigzag_BA:RK} and Fig.~\ref{fig:zigzag_BA:AM}) also support our hypothesis (\ref{rq:engagement}) that students adopting a performance-avoidance strategy could have a measurable impact on the estimation of the transfer ability of the student population using performance data from OLMs.
To mitigate the impact of such strategic guessing as much as possible, we will limit ourselves to studying the 0-1 BA group for both the 2017 and 2018 student sample in the following analysis, for which we are confident that few if any students adopted the performance-avoidance strategy, and the measurements will be accurate. 
It is possible that the other two BA groups, especially the 2-3 BA group, also contain students who frequently guessed due to other reasons such as lack of confidence. 
However, as discussed in section \ref{sec:discussion}, those student are less likely to be the majority in the other two BA groups, and that our current analysis cannot distinguish them from those who guessed because of a performance-avoidance strategy.

We compared the pass rates of the 0-1 BA group from 2018 on modules 2-5 with a propensity score matched subsample in 2017 who also had 0-1 BAs on the first two modules.
The pass-rates for both sequences are shown in Fig.~\ref{fig:zigzag_matched_all}, while the $p$-values from Fisher's exact test comparing each pair of data points on the figures is listed in the first two rows of Table~\ref{tab:zigzag_p_adj}.
All $p$-values are adjusted for Type I error due to conducting multiple tests using the Benjamini-Hochberg method~\cite{benjamini1995controlling}.
The data shows that there are significant performance differences in the success rate between the two student populations on Tutorial Pre and Example 1 Pre attempts in the rotational kinematics sequence, whereas the difference in the angular momentum sequence is less prominent, possibly due to the success rate being very high in both samples.
The differences are similar in nature but larger in magnitude compared to what was observed in our earlier study that did not consider alternative learning strategies~\cite{chen2019perc}, suggesting that the earlier study could have underestimated the transfer ability of the student population due to some students adopting performance-avoidance goals.

\begin{table}[b]
  \caption{A list of $p$-values from Fisher's exact test comparing the performance of 2018 students and matched 2017 students on each common assessment in the listed figure.
  The $p$-values have been adjusted using the Benjamini-Hochberg method~\cite{benjamini1995controlling}. Significant ($p<0.05$) and highly significant ($p < 0.01$) values are marked using * and ** respectively.
  }
  \label{tab:zigzag_p_adj}
  \centering
  \begin{ruledtabular}
  \begin{tabular}{c | c c c c c}
        & Tutorial & Tutorial & Example 1  & Example 1  & \multirow{2}{*}{Quiz} \\
    Fig.  & Pre    & Post   & Pre    & Post   & \\
    \hline
    \ref{fig:zigzag_matched_all:RK}
        & 0.003**   & 0.330   & $<0.001$** & $<0.001$**   & 0.054 \\
    \ref{fig:zigzag_matched_all:AM}
        & 0.333   & 0.265   & 0.166   & 0.306   & 0.166 \\
    \ref{fig:zigzag_matched_onramp:RK_passInPre}
        & 0.001**   & 1.000   & 0.001**   & 0.395   & 0.438 \\
    \ref{fig:zigzag_matched_onramp:RK_passInPost}
        & 0.498   & 1.000   & 0.008**   & 0.028*   & 0.028* \\
    \ref{fig:zigzag_matched_onramp:AM_passInPre}
        & 0.764   & 0.766   & 0.267   & 1.000   & 0.766 \\
    \ref{fig:zigzag_matched_onramp:AM_passInPost}
        & 0.835   & 0.835   & 0.835   & 0.835   & 0.835
  \end{tabular}
  \end{ruledtabular}
\end{table}

\begin{figure*}
  \subfloat[Rotational Kinematics: Matching 2018 Pass On-Ramp Pre students. \label{fig:zigzag_matched_onramp:RK_passInPre}]{\includegraphics[width=0.49\linewidth]{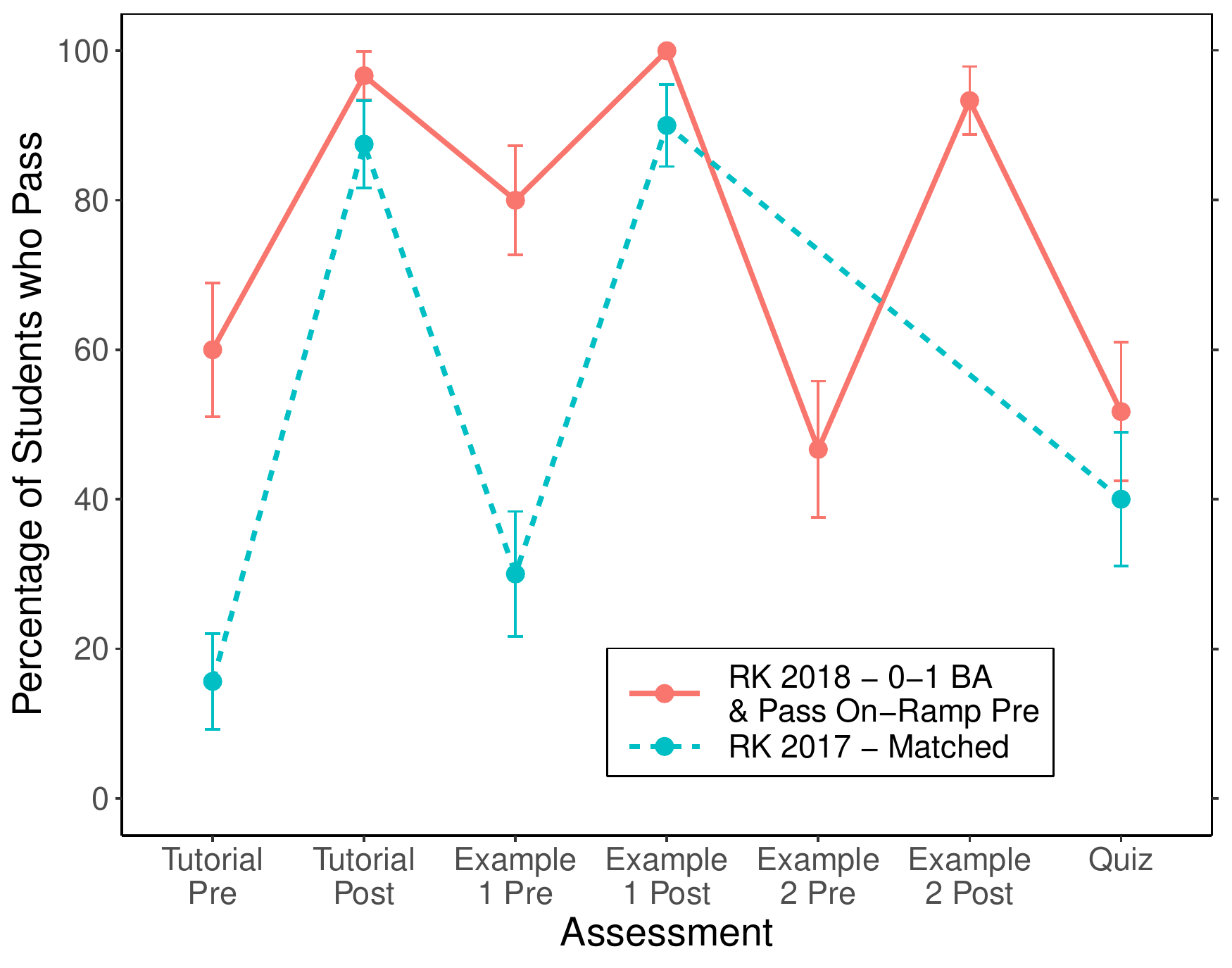}}%
  \hfill
  \subfloat[Rotational Kinematics: Matching 2018 Pass On-Ramp Post students. \label{fig:zigzag_matched_onramp:RK_passInPost}]{\includegraphics[width=0.49\linewidth]{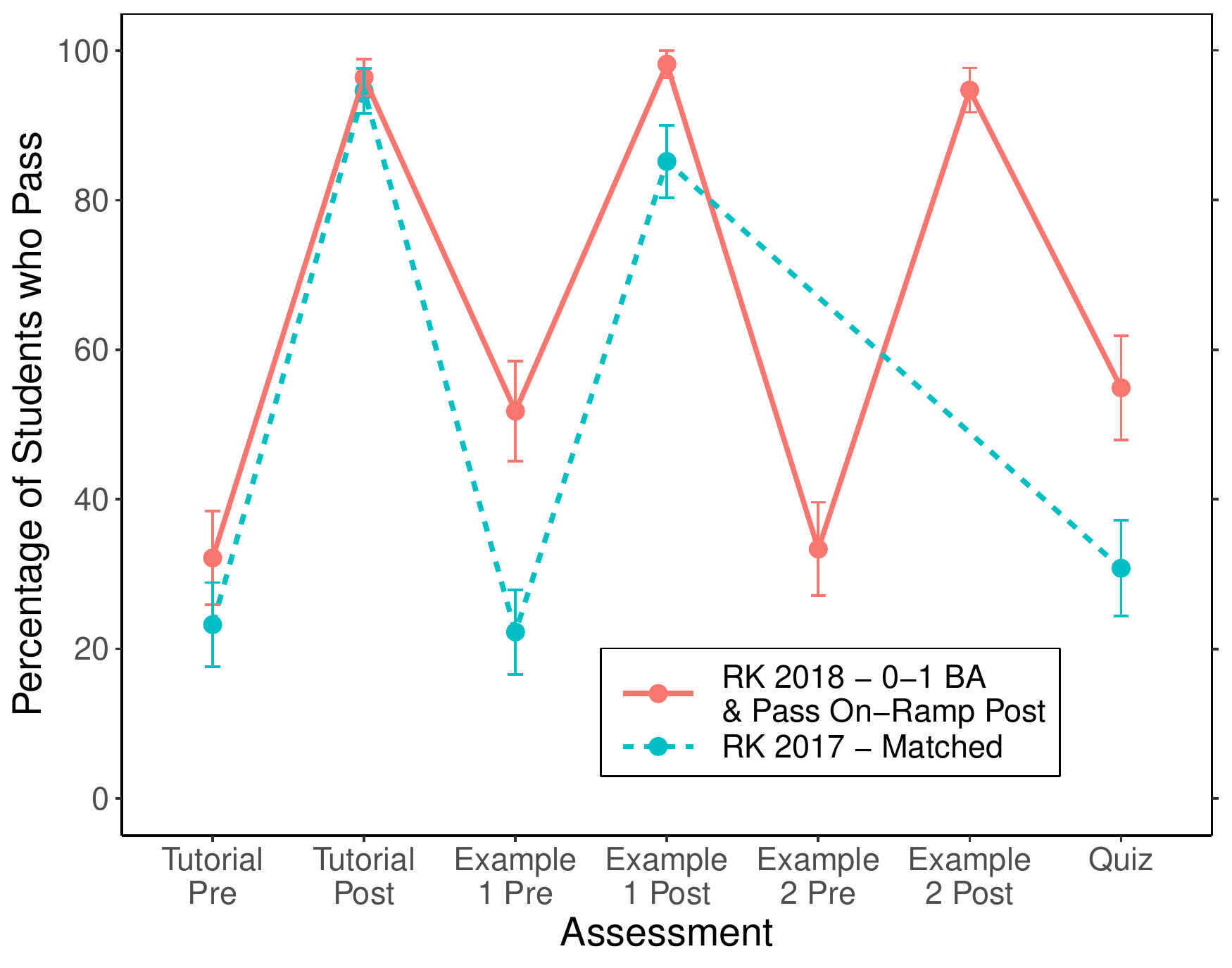}}
  
  \subfloat[Angular Momentum: Matching 2018 Pass On-Ramp Pre students. \label{fig:zigzag_matched_onramp:AM_passInPre}]{\includegraphics[width=0.49\linewidth]{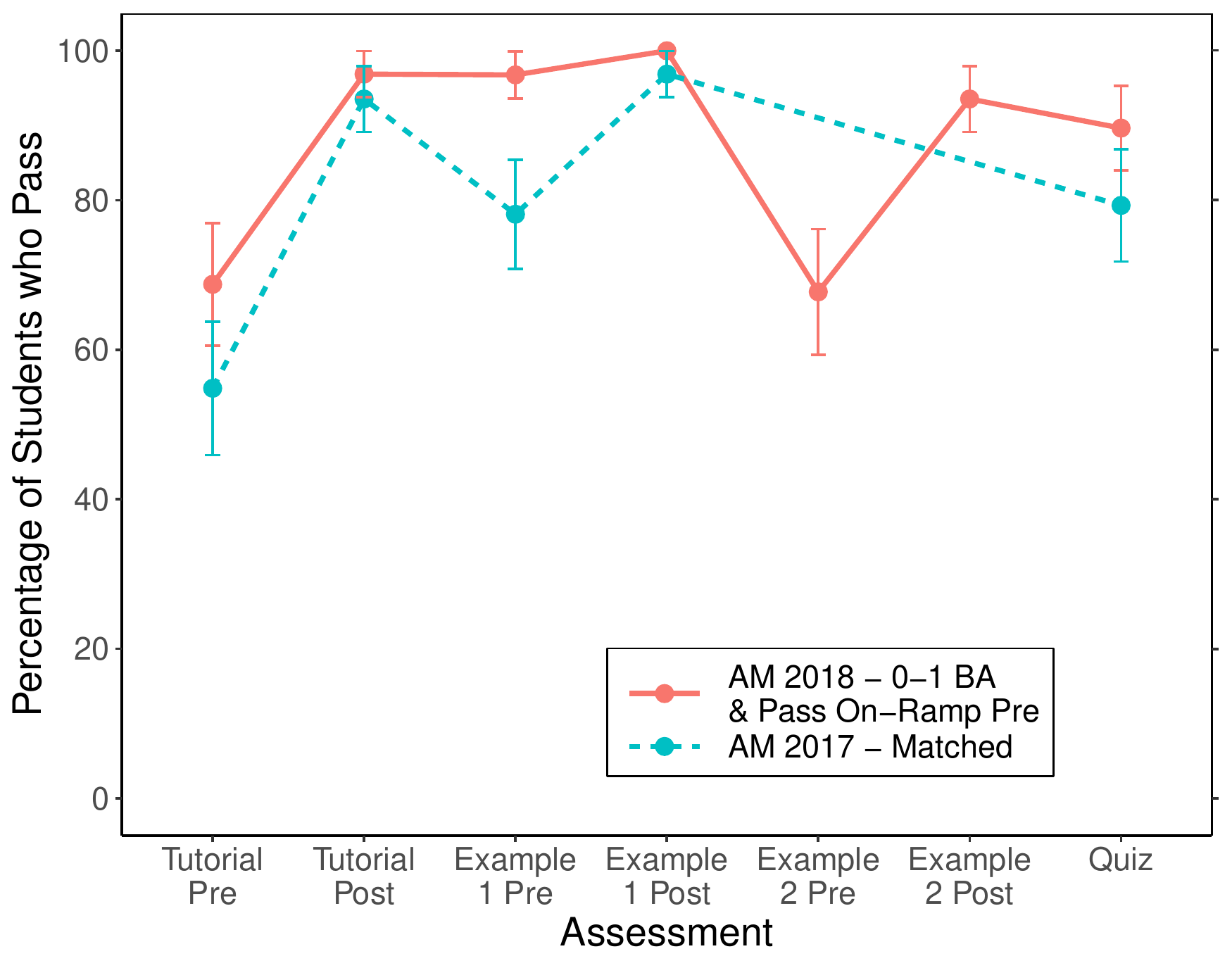}}%
  \hfill
  \subfloat[Angular Momentum: Matching 2018 Pass On-Ramp Post students. \label{fig:zigzag_matched_onramp:AM_passInPost}]{\includegraphics[width=0.49\linewidth]{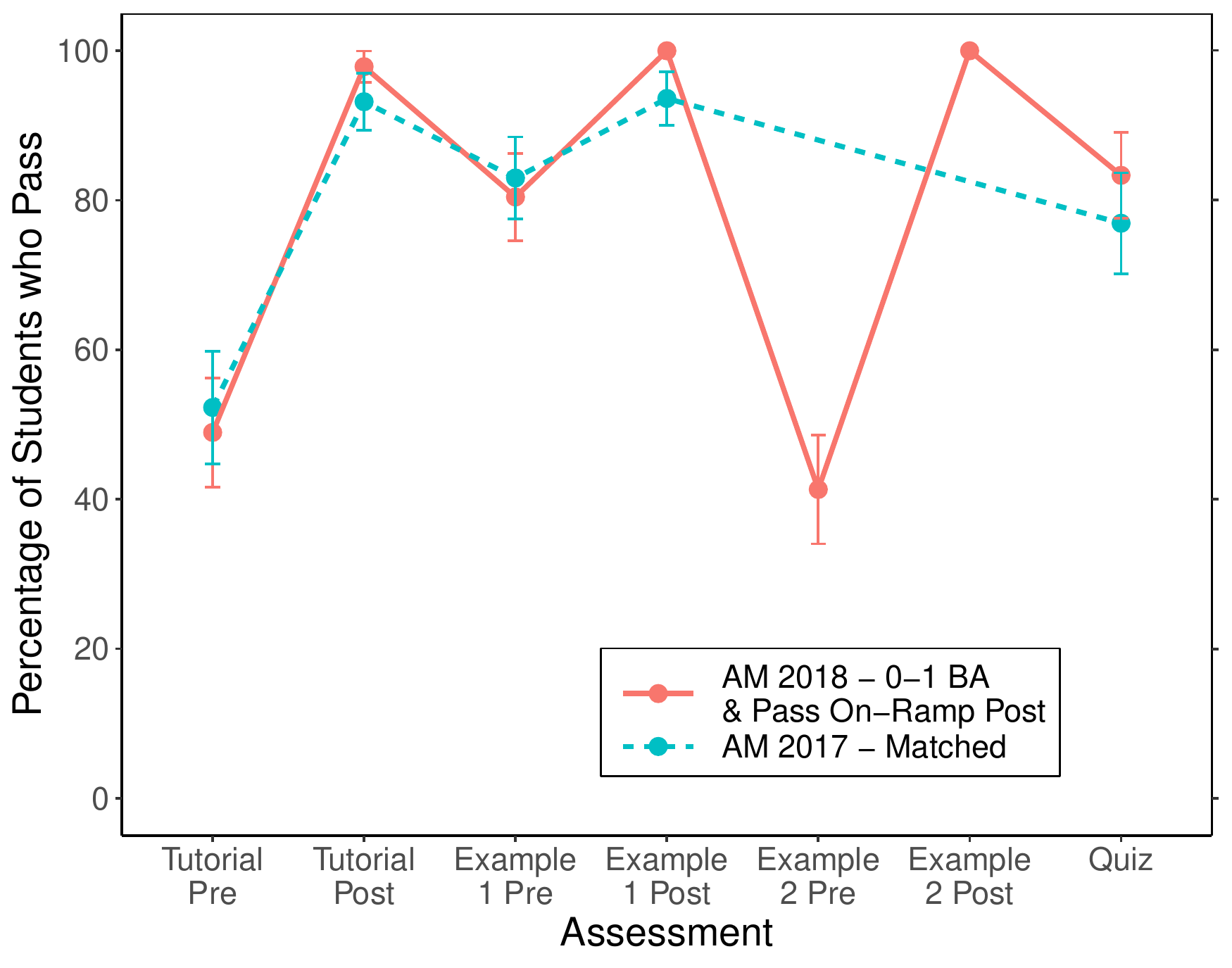}}
  \caption{
  Comparison of the performance on the pre and post attempts of modules for students with 0-1 Brief-attempts and different on-ramp performance a) and c): Pass On Ramp Pre. b) and d): Pass On Ramp Post. 
  The student population in 2017 students were selected to match the background knowledge level of students in 2018 using a propensity score derived from exam scores. Passing rates from the on-ramp modules are not shown.
  Data on Example 2 from 2017 is absent because the module was added in 2018.}
  \label{fig:zigzag_matched_onramp}
\end{figure*}

To examine the mechanism by which the on-ramp module improves the transfer of knowledge (\ref{rq:onramp}), we divided the 2018 0-1 BA population into three cohorts, the number of students in each cohort is listed in Table~\ref{tab:2018_onramp_groups} for each OLM Sequence.
Since the Fail cohort is much smaller than the other two cohorts and too small for reliable propensity score matching, we will only analyze the Pass On-Ramp Pre and Pass On-Ramp Post cohorts (see Table~\ref{tab:2018_onramp_groups}).
In Fig.~\ref{fig:zigzag_matched_onramp}, we compare the performance of those two cohorts to their counterparts in the Fall 2017 semester, using propensity score matching to select a group with similar overall physics ability. The pass rates of the two cohorts on the same module sequence are shown side by side.
Data from the RK sequence is shown on the top row (Fig.~\ref{fig:zigzag_matched_onramp:RK_passInPre} and Fig.~\ref{fig:zigzag_matched_onramp:RK_passInPost}) and the AM sequence in the bottom row (Fig.~\ref{fig:zigzag_matched_onramp:AM_passInPre} and Fig.~\ref{fig:zigzag_matched_onramp:AM_passInPost}).
The adjusted $p$-values of Fisher's exact test between each pair of points are listed in the last four rows of Table~\ref{tab:zigzag_p_adj}.

It can be seen from Fig.~\ref{fig:zigzag_matched_onramp} that the Pass On-Ramp Pre cohort is responsible for the majority of the differences on Pre-study attempts between the 2017 and 2018 samples for the RK sequence, since none of differences are statistically significant for the Pass On-Ramp Post cohort after $p$-value adjustment.
For the AM sequence, neither cohorts showed any statistically significant differences after $p$-value adjustment.
It should be emphasized that these results are valid only for the 0-1 BA group, which clearly did not display a strategic-guessing behavior. 

\section{Discussion}\label{sec:discussion}

\subsection{Interpretation of results}\label{subsec:interpretation-of-results}

We found that roughly half of the students frequently or consistently made abnormally short submissions on their required first attempts on some or all of the first four modules, probably by either guessing or copying the answer from a peer. 
While students may submit an occasional brief attempt due to many reasons, such as lack of self-confidence, we believe that repeated brief attempts are more likely a strategic choice because of the follow reasons. 
First, 35 seconds is barely enough time to carefully read the problem texts. 
There is no clear reason why students who lack confidence would repeatedly and consistently submit answers in such a short amount of time. 
A more likely interpretation is that many of those students are trying to save time. 
Second, failing the first attempt will unlock the instructional materials that significantly boost students' chances of success on the assessment with no grade penalty, providing a good incentive for students to guess without thinking about the problem on their first attempt. 
Third, there were no significant performance differences between the 2-3 brief attempt and 4 brief attempt groups, but a significant difference between the 0-1 brief attempt group and the other two groups, indicating that the 2-3 brief attempt group is more similar to the 4 brief attempt group than the 0-1 brief attempt group.
Finally, there were no detectable performance differences between any of the three brief attempt groups on attempts after studying the learning material, suggesting that the lower performance on initial attempts is less likely due to lower ability level, and more likely the result of strategic random guessing. 
Because of these reasons, we believe that many students with two or more brief attempts likely did so out of a performance-avoidance strategy, which fits well with Boekaerts's description of students being in a ``coping mode''~\cite{boekaerts2000}.
For those students, their goal is to pass the module while saving time and ``unnecessary'' possible failures.

However, it is also worth noting that completely determining the motivation behind student behavior is very difficult by analyzing clickstream data alone. 
While the current analysis shows that students with 0 to 1 brief attempts are less likely to adopt a performance-avoidance strategy, future studies utilizing additional sources of data such as survey and interview will be needed to better estimate how many students did actually adopt such a strategy.

If a student chose to adopt the performance-avoidance strategy, their transfer ability can no longer be measured using OLMs, since their brief Pre-study attempts on the following modules do not always reflect their true ability to transfer their learning from the current module.
If in fact many students in the 2-3 BA and 4 BA group adopted such a strategy, then our follow up analysis including data from those students resulted in an underestimation of students' ability to transfer knowledge from the Tutorial module (module 2) to the Example 1 module (module 3) in our earlier study, although most of the qualitative conclusions remain the same.

An alternative interpretation is that students who frequently adopt the strategy have a lower level of overall mastery on the subject, and possibly a higher level of self-awareness of their lack of knowledge.
Therefore, they would not have been able to pass the required Pre stage attempt even if they had tried, and thus including those students would not under-estimate students' transfer ability.
However, this interpretation seems less likely because students in the 2-3 brief attempt and 4 brief attempt groups performed similarly to the 0-1 brief attempt group on the Tutorial, Example 2 and Quiz modules, as well as on the Post stage of the Example 1 module, which suggests that their overall physics abilities are similar and therefore the observed differences are more likely due to difference in strategical choice.
It must be pointed out that the fact that we excluded almost half of the students in our analysis is by no means an indication that the OLMs are a problematic means of assessment. 
In fact, the average student is likely to adopt avoidance oriented goals on any type of assessment, especially on not for credit assessments such as pre-post conceptual surveys. 
The ability of our current method to estimate the prevalence of such strategies in the student population is actually an advantage over analysis schemes based on traditional paper and pencil tests, or even some earlier studies of online problem solving such as in Ref.~\cite{chenpritchardsingh}, that did not take into account the impact of different student strategies.

Given those results, the current OLM design can provide an accurate measurement of the transfer ability for the sub-population of students who did not frequently make brief submissions on their initial attempts, and an upper bound for the transfer ability for those who did.
For the latter population, more research is needed to determine whether most students did engage in strategic guessing on their first attempt.
If that is indeed the case, then an improved instructional design that discourages such behavior will be needed to more accurately measure their transfer ability.

Another major finding of the current analysis is that, for the remaining students who did not frequently guess on the first attempts, the benefit of the on-ramp module in facilitating transfer (as measured by Pre stage attempts of subsequent modules) predominantly occurs among students who can pass the on-ramp module before accessing the instructional component.
In other words, some students' ability to transfer on subsequent modules were improved by simply doing and passing the problems in the on-ramp module. 
The difference is much more prominent for the more challenging rotational kinematics sequence, and less so for the easier angular momentum sequence.
This observation holds true even after we used propensity score matching between the two semesters to control for the possibility that the Pass On-Ramp Pre cohort could include students with better overall physics knowledge or higher motivation than students in the Pass On-Ramp Post cohort.

A possible explanation could come from the basic principles of information processing theory~\cite{sweller2011, simon1978}.
For students who already possess the essential skills or procedures, attempting the on-ramp module assessment prompted them to retrieve those skills from long-term memory and retain them in working memory.
All or part of those skills remained either in the working memory or in a more active state when those students moved on to the subsequent modules, thereby freeing up cognitive capacity for them to better comprehend the additional complexity of the Tutorial and Example 1 modules.
On the other hand, for those who had not yet mastered those essential skills, the IC of the on-ramp module was sufficient for them to pass the assessment on the next attempt, but not enough for them to achieve a higher level of proficiency.
Therefore, activating those skills on the subsequent modules required a higher amount of cognitive load, limiting students' abilities to process the additional complexities. 

A straightforward and testable implication of this explanation is that providing students with more practice opportunities on those essential skills will increase the ability to transfer on subsequent modules for students with a less solid grasp on those basic skills.
In addition, it may be beneficial to distribute those practices rather than clustering them immediately prior to the tutorial sequence, as distributed practice has been shown to be beneficial for skill acquisition and recall~\cite{dunlosky2013, henderson2015}, and practices of distributed retrieval of factual knowledge have been shown to improve students' physics exam scores~\cite{Gjerde2020}.
It is also worth noting that the significant benefit of having the on-ramp module did not extend to the last quiz module, despite also having an additional Example 2 module in 2018. 
It is possible that additional modules that practice additional basic skills are needed for students to transfer their learning to the last two modules, as they are more complicated and require more skills than was covered in the current on-ramp module. 
Additionally, future studies are needed to apply the same design to other modules or even other courses to examine whether the current results are generalizable across different topics or different disciplines.

Finally, it must be pointed out that our use of propensity score matching to control for the fact that our selected student populations likely have different knowledge and motivation than the rest of the population is far from perfect, since overall exam scores may not fully reflect knowledge on the specific topic involved.
A more accurate propensity score could be constructed in the future, when additional modules on the same topic are created and assigned to students prior to the tutorial sequence.
Such modules have been created and administered in the Fall 2019 semester, enabling more accurate analysis to be conducted in the future.

\subsection{Implications for Online Education Research}\label{subsec:implications}

Our analysis shows that students' behaviors in a self-regulated online learning environment frequently deviate from what was intended or expected by the instructor.
Those behaviors, such as frequently guessing (or cheating in some cases) on problems, could have a substantial impact on the accuracy of assessment and data analysis if not properly accounted for.
Therefore, the ability to detect the presence of diverse student behavior, and account for their potential impact on outcomes of data analysis is a significant advantage of the current OLM based method over conventional assessments such as paper on pencil tests, since students are equally likely to adopt a variety strategies in both situations, yet conventional assessments provide significantly less data on student behavior.

The results of the current analysis can also highlight the importance of future developments in instructional strategies to reduce performance-avoidance strategies among students in an online environment.
In particular, future studies could explore different designs to encourage students to take their first attempts more seriously, such as giving a little bit of extra credit for passing, or do not explicitly reduce the number of attempts after the first try to reduce the perceived cost of attempting to solve the problem. 

Furthermore, in our earlier analysis~\cite{chen2019perc} on the same module sequences, we found that instructional resources designed based on well-documented learning science principles may not always generate expected outcomes due to variations in the actual implementation.
The current analysis further reveals that even when the instructional resource did result in the expected outcome improvement, the underlying mechanism may be different from what was expected.
In this case, modules that were designed to train the proficiency of essential skills among students actually benefited those who were already proficient and did not go through the training by serving as a reminder to activate those skills.
Those results demonstrate the high level of complexity and unpredictability involved in designing and creating effective instructional resources.
Moreover, they highlight the importance of discipline-based education researchers' role as ``Education Engineers'' who bridge the gap between learning theories and actual instructional practices by applying and testing the same design on different content areas and different disciplines.   

Last but not least, the current study is an exploratory attempt at evaluating the effectiveness of instructional materials by comparing the outcomes of students enrolled in two consecutive semesters and controlling for the extrinsic variances using propensity score matching.
Compared to the more common method of conducting randomized AB experiments~\cite{chen2014, chen2016}, the current method is significantly easier to implement in actual classroom settings and introduces fewer disruptions for students compared to randomized control experiments.
In addition, this method allows for a larger sample size since each group consists of an entire class rather than a fraction of the class.
While it introduces more variances due to the treatment and control groups coming from different semesters, we demonstrated that the impact from those variances could be controlled to some extent by methods such as propensity score matching. 
Even though AB testing can provide more rigorous control over extraneous variables, the current setup is far less disruptive to classroom instruction and can be particularly valuable under certain situations, such as during the current COVID-19 pandemic which presents students with many obstacles as institutions shift to fully remote instruction, and instructors are reluctant to introduce more potential sources of confusion.

\acknowledgments{The authors would like to thank the Learning Systems and Technology team at UCF for developing the Obojobo platform. 
Dr. Michelle Taub provided critical and insightful comments on students' self-regulated learning.
This research is partly supported by NSF Grants DUE-1845436 and DUE-1524575 and the Alfred P. Sloan Foundation Grant G-2018-11183.}

\bibliography{main}

\end{document}